\documentstyle[twocolumn,aps,epsf]{revtex}

\begin{document}
\draft                           



\title{Anomalous low-field classical magnetoresistance in two 
dimensions}                          

\author{Alexander Dmitriev$^{1,2}$, Michel Dyakonov,$^1$ and R\'emi 
Jullien$^3$} 

\address{$^1$Laboratoire de Physique Math\'ematique$^\dagger$,  
Universit\'e
         Montpellier 2,
         place E. Bataillon, 34095 Montpellier, France\\
         $^2$A. F. Ioffe Physico-Technical Institute, 194021 St. 
Petersburg,
         Russia\\
         $^3$Laboratoire des Verres$^\dagger$,  Universit\'e Montpellier 2,
         place E. Bataillon, 34095 Montpellier, France\\
$^\dagger$ Laboratoire associ\'e au Centre National de la Recherche
Scientifique (CNRS, France).}


\maketitle


\begin{abstract}
The magnetoresistance of classical two-dimensional electrons scattered 
by randomly distributed impurities is investigated by numerical 
simulation. At low magnetic fields we find for the first time a 
negative magnetoresistance proportional to $|B|$. This 
unexpected behavior is shown to be due to a memory effect specific for 
backscattering events, which was not considered previously.

\end{abstract}


------------------------------------------------------------------

The problem of magnetoresistance in metals and semiconductors has a long 
history, and a large amount of work, both experimental and theoretical, 
was devoted to this subject. Theoretically, the simplest situation is 
that of two-dimensional degenerate non-interacting electrons moving in a 
plane perpendicular to the magnetic field and scattered by static 
impurity potential (the 2D electron gas at low temperatures is also the 
situation for which most of the experiments were done). In this case the 
relevant electron energy is equal to the Fermi energy, and the 
conventional Boltzmann-Drude approach yields zero magnetoresistance, 
i.e. the longitudinal resistivity, $\rho_{xx}$, does not depend on 
magnetic field. Thus the origin of the observed magnetoresistance should 
be looked for beyond the Boltzmann theory.

Most of the research in this domain was focused on the low-field 
magnetoresistance arising from quantum interference effects (weak 
localization) \cite{Larkin}.  Although the pioneering work of of Baskin 
{\it et al} \cite{baskin} long ago demonstrated the importance of 
non-Boltzmann classical memory effects in magnetotransport (see also 
Refs. \cite{polyakov,boby}), only in recent years it was fully 
recognized that there are not only quantum, but also purely classical 
reasons for magnetoresistance, 
\cite{bobynew,basknew,shklovskii,mirlin,dmitriev}, which may be either negative or positive depending on 
the type of impurity potential. Whatever is the case, the classical 
magnetoresistance appears as a consequence of memory effects which are 
beyond the Boltzmann approach.

For the case of strong short-range scatterers, with which we are 
concerned, there is a large negative magnetoresistance at $\beta\ 
$\raise -1.2mm\hbox{$\buildrel>\over\sim$}$\ 1$, 
\cite{bobynew,basknew,dmitriev} where  $\beta 
=\omega_{c}\tau$, $\omega_{c}$ is the electron cyclotron frequency, and  
$\tau$ is the momentum relaxation time. This is related to the existence 
of "circling" electrons,\cite{baskin,boby} which never collide 
with the short-range scattering centers, the fraction of such electrons 
being
$$P=\exp(-2 \pi R/\ell)=\exp(-2 \pi /\beta),\eqno{(1)}$$
where $R=v/\omega_{c}$ is the cyclotron radius, $v$ is the electron 
(Fermi) velocity, and $\ell=v\tau=(Nd)^{-1}$ is the electron mean free 
path.  For the case of isotropic scattering, the corresponding 
magnetoresistance is given by the simple formula (valid with an accuracy better than 2\%): \cite{basknew,dmitriev}
$$\rho_{xx}=\rho_{0}(1-P),\eqno{(2)}$$
where $\rho_{0}$ is the zero-field resistivity. This formula is in good 
agreement with numerical simulation \cite{dmitriev} in the limit 
$d/\ell=Nd^{2}\rightarrow0$, where $d$ is the effective diameter of the 
scattering centers and $N$ is their concentration.  Eq. (2), which is 
valid in this limit, predicts an exponentially small magnetoresistance 
at low fields, when $\beta<<1$. However for finite values of the 
parameter $d/\ell$ a parabolic negative magnetoresistance was found at 
$\beta<<1$.\cite{dmitriev}

The purpose of this Letter is to report a numerical study of the 
low-field classical magnetoresistance of a 2D electron gas with 
short-range scattering centers. Unexpectedly, our numerical simulation 
reveals a new characteristic magnetic field defined by the relation 
$\beta=d/\ell<<1$. For $\beta<<d/\ell$ the (negative) magnetoresistance 
is linear in $B$, while outside this interval, for $d/\ell<<\beta<<1$ 
the dependence is parabolic. We explain the physics of this previously 
unknown phenomenon and we discuss its relevance to the experimental 
results for magnetotransport in random antidot arrays. \cite{dot1,dot2}

In our simulation a point particle (electron) with a given absolute 
value of its velocity, $v$, is scattered by disks of diameter $d$ 
randomly positioned on a plane inside a square box with periodic 
boundary conditions (the box size is a thousand disk diameters). 
Both the hard-disk (Lorentz) model, which exhibits 
anisotropic scattering, and a modified model with isotropic scattering 
are studied. To characterize the coverage, we introduce a dimensionless 
concentration of scatterers $c=\pi Nd^{2}/4<<1$. 

The simulation procedure 
is identical with the one described in Ref. \cite{dmitriev} : an initial
electron position is chosen at random with an initial velocity along
the $x$ direction, the successive velocity directions are determined after each
collision and the conductivity tensor is determined by calculating
the integral of the velocity-velocity correlation function over a time
of $t = 50\tau$. For each value of field and disk concentration
we take an average over 10$^2$ independent disk configurations and,
for each configuration, over 10$^7$ independent trials for the initial
electron position (10 time more than in ref. \cite{dmitriev}).
 
Our main result is presented in Fig. 1, which shows a characteristic 
anomaly in magnetoresistance, $\Delta\rho=\rho_{xx}(B)-\rho_{0}$, 
around zero magnetic field followed by the anticipated parabolic 
dependence on $\beta$ (hard disk model). The anomaly is more visible 
when $\Delta\rho$ is plotted against $\beta^{2}$, see inset. For still 
higher fields the curve follows Eq. 2, as described in Ref. 
\cite{dmitriev}. Similar results were obtained for isotropic scattering 
and other values of $c$.\\



\begin{figure}

\epsfxsize=220pt {\epsffile{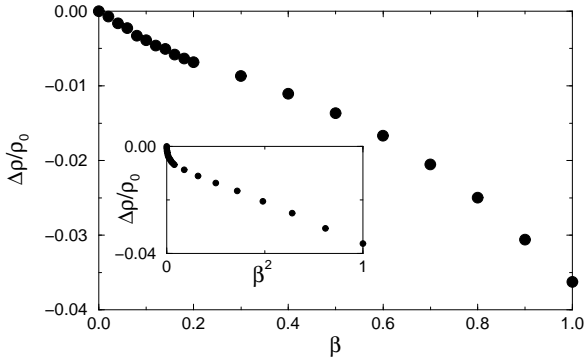}}

\caption{Numerical results for the magnetoresistance 
$\Delta\rho/\rho_{0}$ as a function of $\beta =\omega_{c}\tau$ for 
$c=0.1$ (hard disk model). Inset: same data plotted as a function of 
$\beta^{2}$. Here and below the error bars are within the symbol size.}
 
\end{figure}

To understand this result we need to discuss in some detail the nature 
of the classical non-Boltzmann corrections to the Drude resistivity. The 
Boltzmann approach is equivalent to a random redistribution of all 
scatterers after each collision. In reality there are memory effects of 
two types: (i) the particle may re-collide with the same impurity and 
(ii) the particle's trajectory may repeatedly pass through a space 
region which is free of impurities (the effect of circling electrons is 
of this type).

Consider first the case $B=0$. The classical corrections to the 
Boltzmann equation due to re-collisions with the same scatterer (i) were 
studied long ago. \cite{dorfman} These processes are responsible for the $1/t^{2}$ tail in the velocity correlation function \cite{tails} and for the 
increase of the resistivity compared to its Drude value. The relative 
correction for small $c$ is proportional to $c\ln(1/c)$ (or 
equivalently, to $(d/\ell)\ln(\ell/d)$), the logarithmic term being due 
to the simplest re-collision process $1\rightarrow2\rightarrow1$, while 
longer return loops give a smaller correction on the order of $c$.

As to the type (ii) processes, their role for $B=0$ was not, to our 
knowledge, well understood so far. Consider a particle which, after going a 
distance $x>>d$ without collisions, experiences backscattering at an 
angle  $\phi=\pi$ and then returns to the initial point (Fig. 2$a$). \cite{note} The 
probability of this round trip of length $2x$ is proportional to 
$\exp(-x/\ell)$, not to  $\exp(-2x/\ell)$, as would suggest the 
conventional approach, since the existence of a free corridor of width 
$d$ allowing the first part of the journey guarantees a collision-less 
return. This is not the case for scattering angles outside the interval 
on the order of $d/x$ around the value  $\phi=\pi$, when the 
probabilities for a free path $x$ before and after collision become 
independent and equal to $\exp(-x/\ell)$ each (Fig. 2$b$).

Since typically $x \sim \ell$, the probability of backscattering in the 
interval $\Delta \phi \sim d/\ell$ around  $\phi=\pi$ is enhanced (the 
reversed velocity is conserved for a longer time), and this should lead 
to an additional increase of resistivity on the order of $d/\ell \sim c$, i.e.  same order of magnitude as the contribution of return loops involving 
two or more intermediate scattering. One can say that the existence of 
a free corridor effectively enhances backscattering in the interval 
$\Delta\phi$, roughly by a factor of 2, thus increasing the transport 
cross-section by an amount $\sim d\Delta\phi$.

We attribute the low field anomaly in Fig. 1 to the influence of 
magnetic field on this effect. In the presence of even a small magnetic 
field the electron trajectories before and after collisions can not any 
more follow the same path (Fig. 2$c$). At high enough fields this kills 
the memory effect and, as a consequence, reduces the resistivity (Fig. 
2$d$). Thus a negative magnetoresistance appears with a characteristic 
magnetic field defined by the relation $\beta=d/\ell<<1$.



\begin{figure}

\hskip .8cm \epsfxsize=200pt {\epsffile{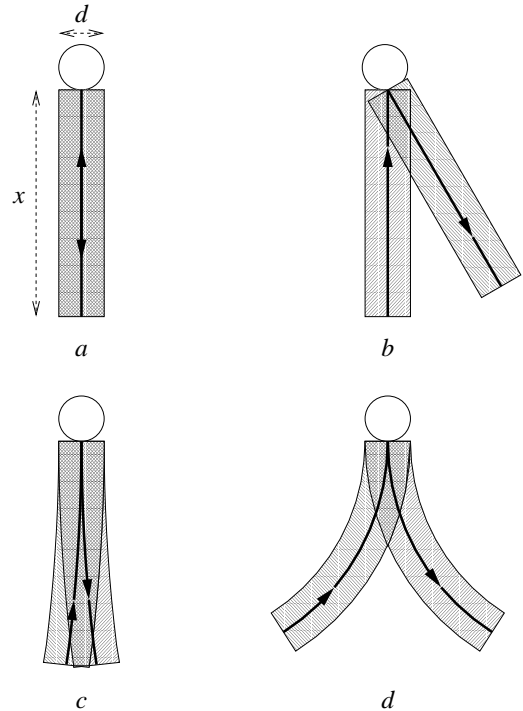}}

\caption{Illustration of the memory effect in backscattering. $a$ - 
Backscattering at angle $\pi$, $B=0$. The particle returns through the 
same corridor. $b$ - Scattering at an angle far from $\pi$, $B=0$. The 
two corridors are different and the probabilities of free flights before 
and after collision are uncorrelated. $c$ - Scattering at angle $\pi$ in 
low magnetic field, $R>>x^{2}/d$. The overlap of the two corridors 
diminishes. $d$ - Same as $c$, but for higher field, $R<<x^{2}/d$. The 
corridors cease to overlap.}
 
\end{figure}

We do not have a regular method for calculating analytically the
magnetoresistance at low fields. However some qualitative conclusions
may be drawn from simple geometric considerations presented in Fig. 2.

Consider again backscattering by an angle exactly equal to $\pi$ but in 
the presence of magnetic field (Fig. 2$c$). In order to have 
collision-less paths of length $x$ before and after scattering, the 
centers of all disks should be outside the corridor of width $d$ 
surrounding these paths. The probability of this is proportional to 
$P=\exp(-NS)$, where $N$ is the disk concentration and $S$ is the joint 
area of the two corridors. The overlapping region should not be counted 
twice.

While at $B=0$ there is full overlap (Fig. 2$a$), $S=xd$ and  
$P=\exp(-x/\ell)$, in the presence of magnetic field the overlap 
diminishes and the relevant area increases. In the low-field limit, this 
increase can be easily calculated to be $\Delta S=x^{3}/(3R) \sim B$, so 
that  $P$ decreases {\it linearly} with $B$. This means that the 
(negative) magnetoresistance is linear in $B$ for $R>>\ell^{2}/d$, or 
$\beta<<d/\ell$.  For higher fields, such that $\beta>>d/\ell$ the two 
corridors practically cease to overlap (Fig. 2$d$) and one has $S=2xd$, 
$P=\exp(-2x/\ell)$. Similar considerations apply to backscattering in 
the interval $\Delta\phi \sim d/\ell$.

A similar contribution comes from the influence of magnetic field on the 
probability of the simplest re-collision process 
$1\rightarrow2\rightarrow1$, which necessarily involves backscattering 
in the same angular interval $\Delta\phi$. At $B=0$ the memory effect 
increases the relative contribution of this process to the resistivity 
by an amount on the order of $c$, and again the curving of the 
trajectories in magnetic field will increase the total area $S$ and thus 
reduce the probability of this process. In the low-field limit one finds 
again that the area increase, $\Delta S$, is linear in $B$. These 
qualitative considerations lead us to the following conclusions 
concerning low-field classical magnetoresistance in two dimensions in 
the presence of appreciable backscattering.

1) A new characteristic magnetic field exists, at which the classical 
parameter $\beta =\omega_{c}\tau$ is small:  $\beta_{c}=d/\ell \sim c<<1$.

2) The total drop of resistivity in the region  $\beta\ $\raise 
-1.2mm\hbox{$\buildrel<\over\sim$}$\ \beta_{c}$ is on the order of 
$d/\ell \sim c$.

3) At $\beta<<\beta_{c}$ the resistivity $\rho_{xx}$ is {\it linear} in 
magnetic field exhibiting the  $|B|$ cusp observed in 
our simulation\cite{fn}.

4) For $\beta_{c}<<\beta<<1$ only quadratic in $B$ corrections remain, 
which are on the order of $c\beta^{2}$. (It can be shown that these 
corrections come from the influence of the magnetic field on the 
contribution of return loops).

This means that at low fields, $\beta<<1$, the magnetoresistance, 
$\Delta\rho$, is described by the formula:
$$\Delta\rho /\rho_{0}=-c(f(\beta /c)+A\beta^{2}),\eqno{(3)}$$
where $A$ is a numerical constant and $f(\xi)$ is a function which 
behaves as $|\xi|$ for small values of its argument and saturates at 
some value on the order of 1 for $|\xi|>>1$. A complete theory should 
give the explicit form of the function $f(\xi)$ and the numerical value 
of the constant $A$. For the hard disk model, we find $f(\xi) \approx 
0.04 |\xi|$ for $\xi\ $\raise -1.2mm\hbox{$\buildrel<\over\sim$}$\ 2$, 
and $A \approx 0.3$.

Note, that Eq. 3 implies that at low field the slope of 
$\Delta\rho/\rho_{0}$ as a function of $\beta$ is independent of $c$ and 
that the derivative $(d/d\beta)(\Delta\rho/\rho_{0})$ has a jump at 
$B=0$ which is a numerical constant depending only on the relative 
efficiency of backscattering ($\approx 0.08$ for hard disks).

Since the conclusions above are not derived rigorously, we test them by plotting the relative magnetoresistance in units of $c$ 
as a function of the dimensionless magnetic field $\beta/c$ for several 
values of $c$. As follows from Eq. 3, in these units the 
magnetoresistance curves should coincide at low fields (where the function  $f(\xi)$ is linear), but diverge  for  $\beta\ $\raise -1.2mm\hbox{$\buildrel>\over\sim$}$\ c$. This prediction is in excellent agreement with the simulation results, as presented in Fig. 3, which convinces us that our understanding is basically correct.\\


\begin{figure}

\epsfxsize=240pt {\epsffile{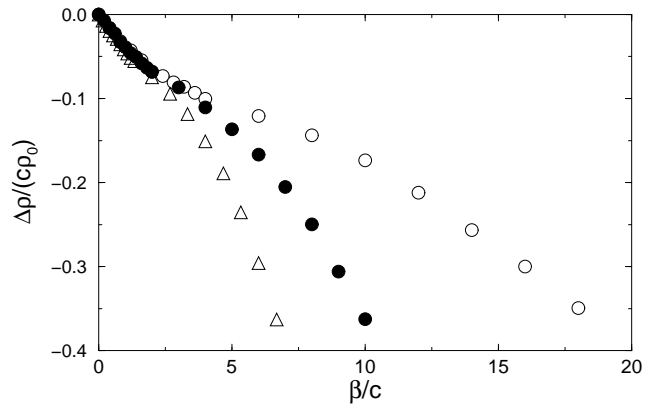}}

\caption{Simulation results for several values of $c$ plotted in units 
$\Delta\rho/(c\rho_{0})$ versus $\beta/c$ showing universal behavior at 
low fields as predicted by Eq. 3. Open circles, filled circles and
open triangles correspond to $c =$ 0.05, 0.1, and 0.15, respectively.}

\end{figure}

Finally, we compare our simulation with the experimental results for 
magnetoresistance of 2D electrons in a disordered array of antidots. 
\cite{dot1,dot2} In these experiments the antidot diameter, $d$, 
is much greater than the Fermi wavelength, so that the classical approach 
is justified. Our numerical simulation exactly corresponds to the experimental 
situation at low temperatures. 

Negative linear magnetoresistance at low fields, first 
observed in Ref. \cite{dot1}, quantitatively agrees with our results as 
seen in Fig. 4. The necessary data being given in Ref. \cite{dot1}, this 
comparison is done without any fitting parameters. (The main uncertainty 
is in the effective antidot diameter which is given in Ref. \cite{dot1} 
as $d=0.2-0.3 \mu m$. We take  $d=0.25 \mu m$. In order to re-plot the 
experimental curve we calculate $c \approx 0.1$ and $\beta=12.5B$(T)). 
It should be noted that while there is an excellent agreement for  
$\beta\ $\raise -1.2mm\hbox{$\buildrel<\over\sim$}$\ 1$, strong 
deviations are found at $\beta>1$. The discrepancy may be due to the 
fact that the antidot array used in these experiments was, in fact, a 
strongly distorted regular lattice with a substantial long-range order.



\begin{figure}

\epsfxsize=220pt {\epsffile{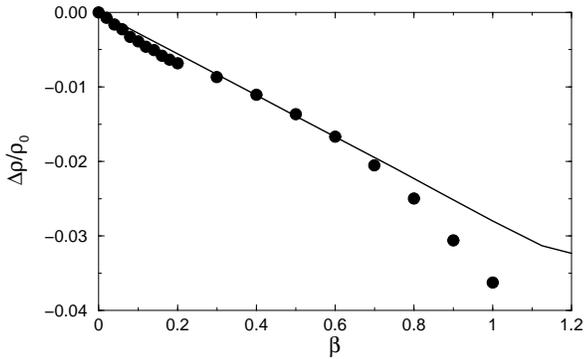}}

\caption{Simulation results for magnetoresistance at $c=0.1$ (filled circles) 
compared to 
the  experimental curve of Ref.[10] (thick line).}

\end{figure}

In Ref. \cite{dot2} the observed anomaly is attributed to the quantum 
(weak localization) effect. Direct comparison of our results with the 
experiment is difficult because of the way in which the data is 
presented, however we find, at least, a qualitative agreement both in the 
magnitude of the low-field anomaly and in the value of the 
characteristic magnetic field at T=1.2 K.  This may be an indication that 
the experimental results are (at least partly) a manifestation of the classical 
phenomenon presented above. A strong temperature dependence of the low-field 
anomaly was observed in Ref. \cite{dot2}, which seems to support
 its interpretation as a result of weak localization. However, the observed shrinking of the 
low-field anomaly, as well as of the value $\rho_{0}$, with increasing 
temperature could be also due to the reduction of the memory effect by 
electron-electron and phonon scattering. The temperature dependence of the 
anomalous magnetoresistance needs further studies.

The artificial random antidot array is an interesting object, 
and a systematic experimental study of magnetoresistance in this system 
is highly desirable to separate the classical and quantum effects in magnetotransport.

A regular analytical calculation of the classical magnetoresistance at 
low fields remains a challenging task. It would also be important to 
understand the transition from classical to quantum behavior as the 
ratio of the Fermi wavelength to the scattering diameter increases.

Interestingly, a mechanism similar to the one considered here is 
responsible for the "opposition effect" \cite{hapke}  - the increase of 
the brightness of the Moon and other planets in a small angular interval 
around the position when the Sun, the planet, and the observer lie on 
the same straight line.

In summary, by numerical simulation we have found an anomalous low-field 
classical magnetoresistance of 2D electrons in the presence of strong 
short-range scattering. A new characteristic magnetic field is found, 
at which the classical parameter $\beta=\omega_{c}\tau$ is small: 
$\beta_{c}=d/\ell<<1$. For $\beta<<\beta_{c}$ the negative magnetoresistance 
is proportional to $|B|$, and the slope of $\Delta\rho/\rho_{0}$ versus $\beta$ 
is independent of the impurity concentration. We have shown that this phenomenon 
is due to a specific memory effect associated with backscattering.

We appreciate useful discussions with D. Polyakov and Z.D. Kvon. 
We thank B. van Tiggelen for drawing our attention to the opposition effect and to Ref. 
\cite{hapke}\\


%


\begin{references}
\bibitem{fn} Strictly speaking, the magnetoresistance {\rm is} an
analytical function of $B$: a careful analysis shows that, in a very
small region around zero field, where 
$\beta\ $\raise -1.2mm\hbox{$\buildrel<\over\sim$}$\ c^2$, the dependence
on $B$ should be parabolic. This region is not accessible in our 
simulation.
\end{references}
\end{document}